\documentclass{article}
\pdfoutput=1

\usepackage[english]{babel}
\usepackage[utf8]{inputenc}
\usepackage[T1]{fontenc}

\usepackage{textgreek}
\usepackage{enumitem}
\usepackage{gensymb} 

\usepackage[a4paper,top=2.5cm,bottom=3cm,left=2.5cm,right=2.5cm,marginparwidth=1.75cm]{geometry}

\usepackage{amsmath}
\usepackage[round]{natbib}
\usepackage{graphicx}
\usepackage{hhline}
\usepackage{longtable}
\usepackage[colorlinks=true, allcolors=blue,hyperfootnotes=true]{hyperref}
\usepackage[detect-weight,retain-explicit-plus]{siunitx} %

\usepackage{tikz}
\usetikzlibrary{shapes,arrows}

\usepackage[yyyymmdd]{datetime}

\title{Introduction to Ambisonics, Part 1:\\
The Part With No Math}

\author{Jens Ahrens}
\date{Technical note\footnote{This document is part of the course \emph{Spatial Audio -- Practical Master Guide}: \url{https://spatial-audio.acoucou.org/}.  \newline \indent \hspace{2.5mm}A video presentation of this document is available at \url{https://www.youtube.com/playlist?list=PLG4Vmg8e9fh1OwX07YP87Aap-c1n692Ur}.}~~v.~1.0\\[1ex] 
Chalmers University of Technology\\[1ex]{jens.ahrens@chalmers.se} 
\texttt{jens.ahrens@chalmers.se}}

\usepackage{natbib,amsmath}
\usepackage{graphicx}
\usepackage{tikz}
\usepackage{tikz-3dplot}

\begin{document}

\maketitle

\begin{abstract}
The present document is Part 1 of a 2-part introduction to ambisonics and aims at readers who would like to work practically with ambisonics. We leave out deep technical details in this part and focus on helping the reader to develop an intuitive understanding of the underlying concept. We explain what ambisonic signals are, how they can be obtained, what manipulations can be applied to them, and how they can be reproduced to a listener. We provide a variety of audio examples that illustrate the matter. Part 2 of this introduction into ambisonics is provided in a separate document and aims at readers who would like to understand the mathematical details.
\end{abstract}

\vspace{10mm}
\noindent (Part 2 of this introduction to ambisonics is in preparation.)

\section{Introduction}

\subsection{What is Ambisonics?}

You might have asked this question before, and the responses that you received were likely to be confusing. This is not a surprise as the one and only definition of ambisonics does not exist. The way one thinks about ambisonics nowadays can be very different from how the pioneers in the 1970s were thinking of it back then. 

The term ambisonics is derived from Latin and may be translated to `surround sound'. Ambisonics is a concept for spatial audio capture, storage, and reproduction that is fundamentally different from any of the other concepts that exist, which may be the reason for why it resides in its own microcosm. It was pioneered in the 1970s~\citep{Fellgett:SS1075,Gerzon:SS1075}, but it did not experience commercial success then. A small number of enthusiasts kept the ball rolling over the decades until the ready availability of multichannel audio hardware and of sufficient digital processing power made the academic research community pick it up again around the early 2000s from which point on the concept matured to the extent that it was finally ready to trickle from academia into practice. Software tools have been available freely and open source for years, and the significance of ambisonics in the industry keeps growing. Around the year 2016, with the rise of 360 videos viewed using VR goggles, even big players like YouTube\textsuperscript{TM} and Facebook\textsuperscript{TM} adopted ambisonics as audio format for their media players. Especially when thinking of spatial audio for VR goggles, it is hardly thinkable that anything serious can be achieved that is not closely related to the ambisonics. 

One can think of ambisonics as a concept that is built around a specific type of multichannel audio signal called an ambisonic signal\footnote{The term \emph{ambisonics signals} (with ``s'') is also common. Native English speakers emphasized to the author that even though the concept is termed \emph{ambisonics}, it is more appropriate to leave out the ``s'' when the term is used as an adjective.}, and it comprises the entire signal pipeline from capture to reproduction. The channels of the ambisonic signal are related to each other through an advanced mathematical framework that describes the physical structure of a sound field. One speaks of the sound field or sound scene being \emph{encoded} into ambisonics. The underlying mathematical framework is indeed rather involved. It bases on functions that are termed \emph{spherical harmonics} or \emph{spherical surface harmonics}. These spherical harmonics are used extensively in a variety of fields of physics including in quantum mechanics. The good news is that it is absolutely not necessary to understand the underlying mathematics to be able to work with ambisonics. We therefore omit diving deep into it here. We refer the interested reader to~\citep{Hollerweger:ambisonics2005,Zotter:book2019,Armstrong:Routledge2022,Ahrens:ambisonics_part_2}.

From a practical point of view, an ambisonic signal is a multichannel audio signal. But because the set of channels constitute a comprehensive representation of an entire audio scene, we will sometimes use the terms `ambisonic representation' or `ambisonic content' or similar to refer to it. One important aspect to understand is the circumstance that an ambisonic representation relates to a given point in space, which is the vantage point from which the sound scene is observed. Here is a quick example that illustrates what this means:

Imagine someone playing a violin in a room. One can place a suitable arrangement of microphones (a so-called microphone array) into the room and obtain an ambisonic representation of the physical structure of the sound field around that microphone array. This means that the ambisonic signal contains information on, for example, from what direction the sound waves that the violin radiates impinge on the microphone array and with what curvature, when and from what direction room reflections impinge with what spectral signature, and the like -- all as observed at the location of the microphone array. 

Doesn't this sound like a very powerful framework? What is a little unfortunate is that the ambisonic representation of the sound field is rather abstract so that one cannot directly read from the signals things like how many sound sources there are in the scene, where they are located, how strong the reverberation is and the like (This is, by the way, a very difficult endeavor with any signal representation.). The powerful aspect of ambisonics is that the sound field that an ambisonic signal represents can be physically re-created (even though we do not know in tangible terms what exactly the sound field contains). In other words, one can drive an array of loudspeakers such that the sound waves that the individual loudspeakers emit superpose and make up the original captured sound field. It is even possible to render the ambisonic signal binaurally, which is equivalent to virtually placing a human head in the sound field and compute the signals that would arise at the ears of the person if they would have been listening to the original scene at the location of the microphone arrangement that captured the ambisonic signal. We will explain this in more detail in Sec.~\ref{sec:reproduction}.

Summarizing the above and attempting a definition of ambisonics, one could say that ambisonics is the combination of a specific multichannel representation of a spatial audio scene, which is based on spherical harmonics, and the corresponding capture and reproduction technologies. Fig.~\ref{fig:ambisonics_pipeline} summarizes this. Before explaining to you how one can capture and reproduce an ambisonic signal, we would like you to obtain a better feel for what ambisonic signals are.

\tikzset{
  Speaker/.pic={
    \filldraw[fill=white,pic actions] 
    (-15pt,0) -- 
      coordinate[midway] (-front) 
    (15pt,0) -- 
    ++([shift={(-6pt,10pt)}]0pt,0pt) coordinate (aux1) -- 
    ++(-18pt,0) coordinate (aux2) 
    -- cycle 
    (aux1) -- ++(0,8pt) -- coordinate[midway] (-back) ++(-18pt,0) -- (aux2);
  },
  Mic/.pic={
    \draw (0,0) circle[radius=3mm];
    \draw (-3mm,3mm) to (3mm,3mm);
    \coordinate (-back) at (0mm,-3mm);
    }
}

\tikzstyle{block} = [draw,fill=gray!20,minimum size=2em]
\tikzstyle{circle} = [draw,shape=circle,fill=gray!20,minimum size=2em]
  
\begin{figure}[tbh]
\begin{center}
\begin{minipage}{.49\textwidth}
\raggedright
\begin{tikzpicture}[>=latex']

    \node at (-2.75, 2.8) {\scriptsize \textcolor{gray}{Microphone}};
    \node at (-2.75, 2.4) (label1) {\scriptsize \textcolor{gray}{signals}};
    \draw[gray,->] (label1) -- (-2.75, 1.5);

    \pic[rotate=90,scale=.4] (mic1) at (-3, 1)  {Mic};
    \pic[rotate=90,scale=.4] (mic2) at (-3, .5)  {Mic};
    \pic[rotate=90,scale=.4] (mic3) at (-3, 0)  {Mic};
    \pic[rotate=90,scale=.4] (mic4) at (-3, -1)  {Mic};

    \node[block, minimum height=3cm] at (-2, 0) (block1) {\rotatebox{90}{Encoding}};
    
    \draw[transform canvas={yshift=1cm},->]  (mic3-back) -- (block1);
    \draw[transform canvas={yshift=0.5cm},->]  (mic3-back) -- (block1);
    \draw[transform canvas={yshift=0.0cm},->]  (mic3-back) -- (block1);
    \draw[transform canvas={yshift=-1cm},->]  (mic3-back) -- (block1);
    \node at (-2.66, -0.38) {\vdots};

    \node at (0, 2.8) {\scriptsize \textcolor{gray}{Ambisonic}};
    \node at (0, 2.4) (label2) {\scriptsize \textcolor{gray}{signal}};
    \draw[gray,->] (label2) -- (0, 1.5);
    
    \node[block, minimum height=3cm] at (2, 0)  (block3) {\rotatebox{90}{Decoding}};
    
    \draw[transform canvas={yshift=1.25cm},->] (block1) -- (block3);
    \draw[transform canvas={yshift=0.75cm},->] (block1) -- (block3);
    \draw[transform canvas={yshift=0.25cm},->] (block1) -- (block3);
    \draw[transform canvas={yshift=-1.25cm},->] (block1) -- (block3);
    
    \node at (0, -.38) {\vdots};

    \node at (2.67, 2.8) {\scriptsize \textcolor{gray}{Loudspeaker}};
    \node at (2.67, 2.4) (label3) {\scriptsize \textcolor{gray}{signals}};
    \draw[gray,->] (label3) -- (2.67, 1.5);
    
    \pic[rotate=90,scale=.4] (speaker1) at (3.3, 1)  {Speaker};
    \pic[rotate=90,scale=.4] (speaker2) at (3.3, .5)  {Speaker};
    \pic[rotate=90,scale=.4] (speaker3) at (3.3, 0)  {Speaker};
    \pic[rotate=90,scale=.4] (speaker4) at (3.3, -1)  {Speaker};
    \node at (2.67, -.38) {\vdots};

    \draw[transform canvas={yshift=1cm},->] (block3) -- (speaker3-back);
    \draw[transform canvas={yshift=0.5cm},->] (block3) -- (speaker3-back);
    \draw[transform canvas={yshift=0.0cm},->] (block3) -- (speaker3-back);
    \draw[transform canvas={yshift=-1cm},->] (block3) -- (speaker3-back);
    
\end{tikzpicture}
\end{minipage}
\begin{minipage}{.49\textwidth}
\raggedleft
\begin{tikzpicture}[>=latex']

    \node at (-2.75, 2.8) {\scriptsize \textcolor{gray}{Microphone}};
    \node at (-2.75, 2.4) (label1) {\scriptsize \textcolor{gray}{signals}};
    \draw[gray,->] (label1) -- (-2.75, 1.5);

    \pic[rotate=90,scale=.4] (mic1) at (-3, 1)  {Mic};
    \pic[rotate=90,scale=.4] (mic2) at (-3, .5)  {Mic};
    \pic[rotate=90,scale=.4] (mic3) at (-3, 0)  {Mic};
    \pic[rotate=90,scale=.4] (mic4) at (-3, -1)  {Mic};

    \node[block, minimum height=3cm] at (-2, 0) (block1) {\rotatebox{90}{Encoding}};
    
    \draw[transform canvas={yshift=1cm},->]  (mic3-back) -- (block1);
    \draw[transform canvas={yshift=0.5cm},->]  (mic3-back) -- (block1);
    \draw[transform canvas={yshift=0.0cm},->]  (mic3-back) -- (block1);
    \draw[transform canvas={yshift=-1cm},->]  (mic3-back) -- (block1);
    \node at (-2.66, -0.38) {\vdots};

    \node at (0, 2.8) {\scriptsize \textcolor{gray}{Ambisonic}};
    \node at (0, 2.4) (label2) {\scriptsize \textcolor{gray}{signals}};
    \draw[gray,->] (label2) -- (.75, 1.5);
    \draw[gray,->] (label2) -- (-.75, 1.5);
    
    \node[block, minimum height=3cm] at (0, 0)  (block2) {\rotatebox{90}{~(Editing)~}};

    \draw[transform canvas={yshift=1.25cm},->] (block1) -- (block2);
    \draw[transform canvas={yshift=0.75cm},->] (block1) -- (block2);
    \draw[transform canvas={yshift=0.25cm},->] (block1) -- (block2);
    \draw[transform canvas={yshift=-1.25cm},->] (block1) -- (block2);
    \node at (-1, -.38) {\vdots};
    
    \node[block, minimum height=3cm] at (2, 0)  (block3) {\rotatebox{90}{Decoding}};

    \draw[transform canvas={yshift=1.25cm},->] (block2) -- (block3);
    \draw[transform canvas={yshift=0.75cm},->] (block2) -- (block3);
    \draw[transform canvas={yshift=0.25cm},->] (block2) -- (block3);
    \draw[transform canvas={yshift=-1.25cm},->] (block2) -- (block3);
    \node at (1, -.38) {\vdots};

    \node at (2.67, 2.8) {\scriptsize \textcolor{gray}{Loudspeaker}};
    \node at (2.67, 2.4) (label3) {\scriptsize \textcolor{gray}{signals}};
    \draw[gray,->] (label3) -- (2.67, 1.5);
    
    \pic[rotate=90,scale=.4] (speaker1) at (3.3, 1)  {Speaker};
    \pic[rotate=90,scale=.4] (speaker2) at (3.3, .5)  {Speaker};
    \pic[rotate=90,scale=.4] (speaker3) at (3.3, 0)  {Speaker};
    \pic[rotate=90,scale=.4] (speaker4) at (3.3, -1)  {Speaker};
    \node at (2.67, -.38) {\vdots};

    \draw[transform canvas={yshift=1cm},->] (block3) -- (speaker3-back);
    \draw[transform canvas={yshift=0.5cm},->] (block3) -- (speaker3-back);
    \draw[transform canvas={yshift=0.0cm},->] (block3) -- (speaker3-back);
    \draw[transform canvas={yshift=-1cm},->] (block3) -- (speaker3-back);
    
\end{tikzpicture}
\end{minipage}
\caption{Two flow charts of the complete signal chain in ambisonics. The right chart includes an optional stage for editing of the sound scene. The editing can relate to both spatial and non-spatial information. \emph{Encoding} is a mathematical process that converts microphone array signals or virtual sound scenes that are computer generated into an ambisonic signal. \emph{Decoding} is a mathematical process that converts an ambisonic signal to loudspeaker or headphone signals. All editing of the sound scene takes place based on the ambisonic signals.}
\label{fig:ambisonics_pipeline}
\end{center}
\end{figure}
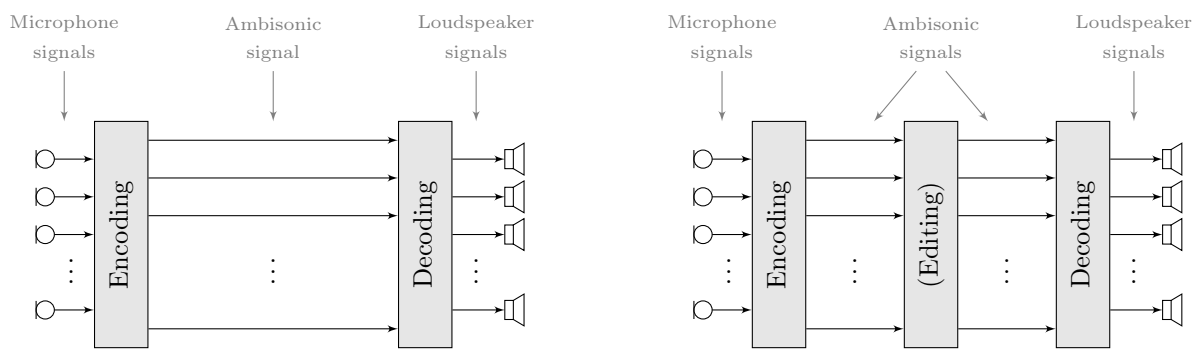
%

\section{Ambisonic Signals}

As stated above, ambisonic signals are multichannel signals. A very fundamental property of an ambisonic signal is its ambisonic \emph{order}, or, simply its order. The order determines the accuracy with which the encoded sound field is represented. A higher order means a more accurate representation. We will discuss the practical significance of the order in Sec.~\ref{sec:ambi_order}. As a ballpark, one could say for now that any order above 5 may be considered high, and any order below that may be considered low. This distinction is not rigid and historically, any order higher than~1 was termed 'high'. 

\begin{figure}[b!]
    \centering
    \includegraphics[width=\columnwidth,trim=-1.3in -.7in -.3in 0in, clip=true]{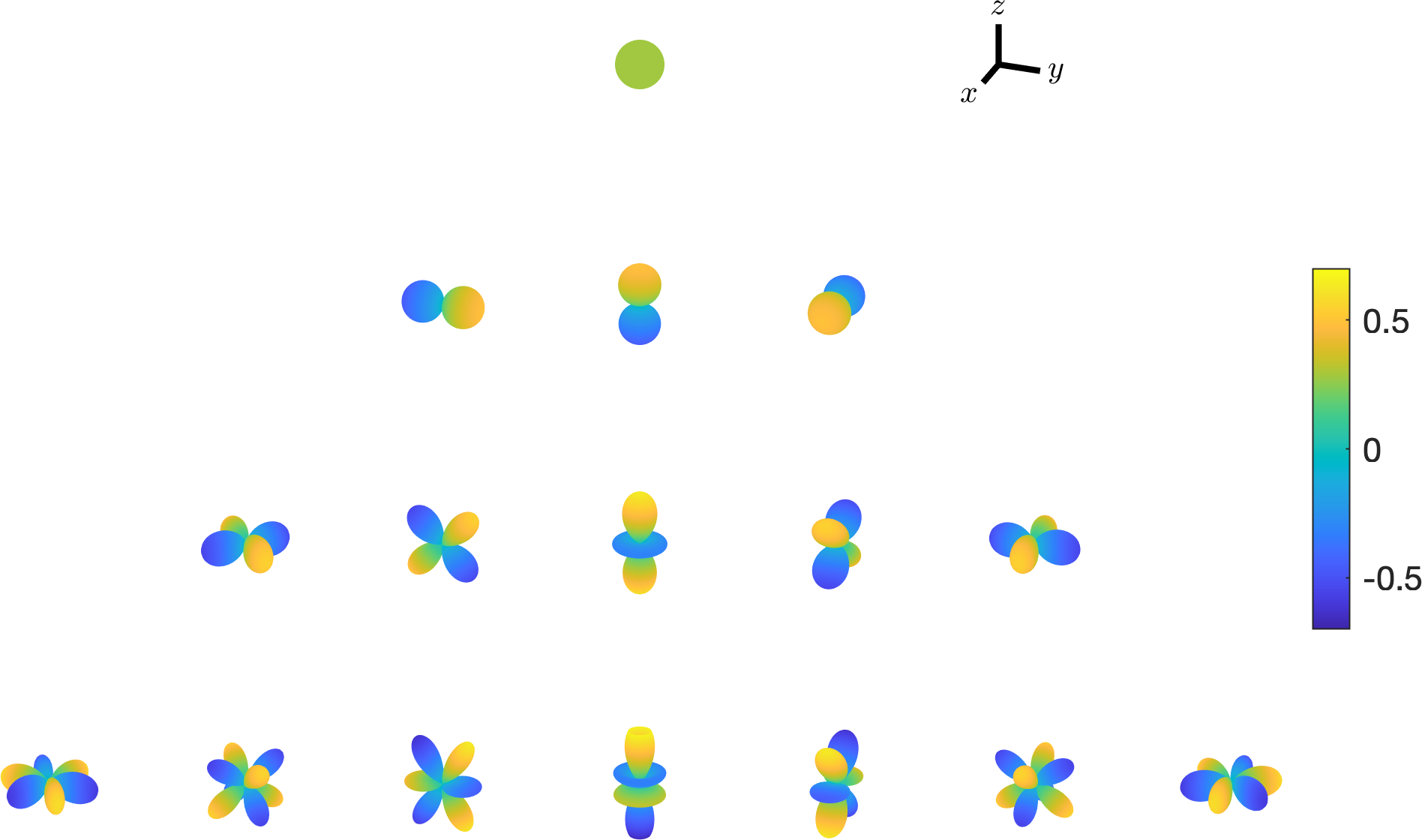}%
    \put(-445, 242){$n=0$}\put(-445, 180){$n=1$}\put(-445, 118){$n=2$}\put(-445, 53){$n=3$}
    \put(-390, 10){$m=-3$}\put(-337, 10){$m=-2$}\put(-282, 10){$m=-1$}\put(-228, 10){$m=0$}\put(-175, 10){$m=1$}\put(-125, 10){$m=2$}\put(-70, 10){$m=3$}
    \caption{Illustration of the spherical harmonics for the orders $n = 0,1,2,3$. These shapes also correspond to the directivities that a theoretical microphone would need to have for recording the given channel of the ambisonic signal. A 0th-order signal comprises only the mode $(n, m) = (0, 0)$. A 1st-order signal (a signal with maximum order of $N = 1$) comprises the modes $(n, m) = (0, 0)$, $(1, -1)$, $(1, 0)$, $(1, 1)$ and so forth. ACN starts at the top and walks through the rows from left to right. The black lines at the top right indicate the right-hand Cartesian coordinate system. The convention is that the $x$-axis points into the direction `straight ahead'.}
    \label{fig:sh_table}
\end{figure}
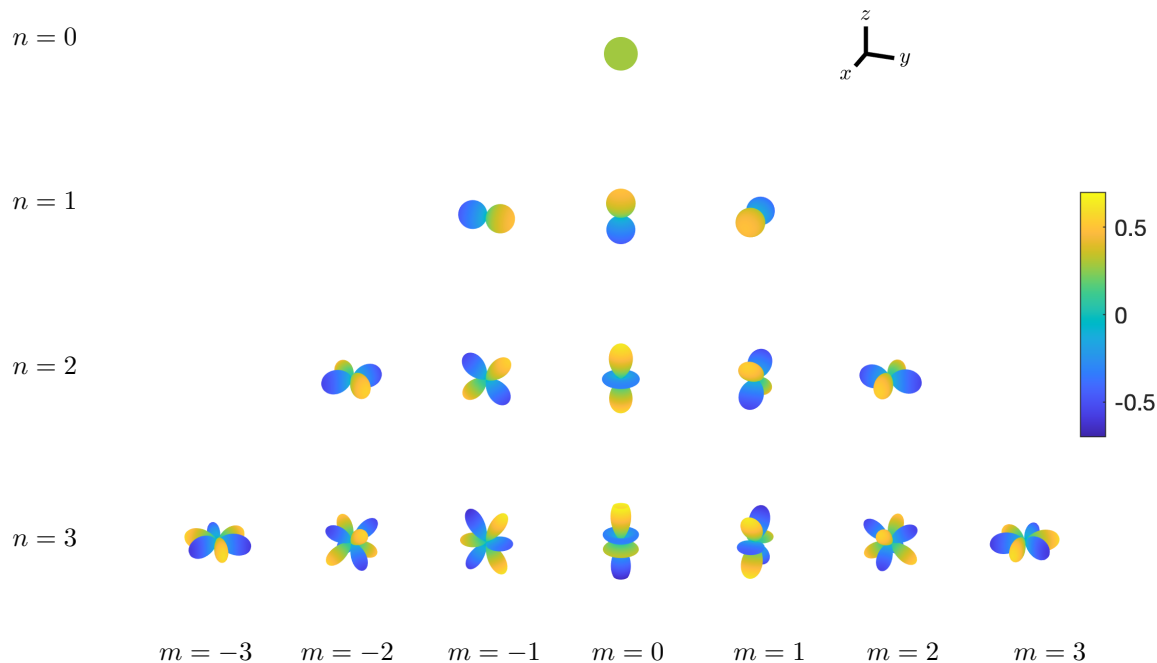

As you might have guessed, a higher-order signal has more channels than a lower order signal. If $N$ is the order, then the ambisonic signal has $(N+1)^2$ channels (a 0th-order signal has one channel, a 1st-order signal has 4 channels, and a 7th-order signal has 64 channels). Usually, the channels in an ambisonic signal are ordered according to the ambisonic channel number (ACN) in which the first channel comprises the 0th-order representation of the signal, the first 4 channels are the 1st-order representation and so forth. Fig.~\ref{fig:sh_table}, which will be introduced shortly, provides more details on the ACN.

\setcounter{footnote}{2}
%


One can think of the first channel of the ambisonic signal to contain the signal that an omnidirectional microphone would capture when being positioned at the vantage point from which the sound scene is observed. Channels~2, 3, and 4 contain signals that are similar to the signals that figure-of-eight microphones would capture at the vantage point when being oriented along the three coordinate axes, respectively. Channels 5 and higher contain signals as they would be captured by microphones with more complicated and detailed directivities. Refer to Fig.~\ref{fig:sh_table} for an illustration of what directivity a single microphone would need to have to capture a signal similar to what is contained in a given channel of an ambisonic signal. We deliberately use the term ``similar'' here because there are some delicate differences between what such notional microphones would capture and what the channels of an ambisonic signal contain. The differences are not of relevance in this context so that we omit going into further detail here. Actual ambisonic signals are not obtained from literal microphones with according directivities but from microphone arrays in which the signals from the different microphones are combined using signal processing to synthesize the required directivity, or they are computer generated. We will cover this in Sec.~\ref{sec:encoding}.

There is one detail of the mathematical framework of spherical harmonics that we would like to introduce here: A given order of an ambisonic representation usually comprises several so-called modes (only the 0th order comprises a single mode). We usually use the symbol $n$ to refer to the order that we are looking at and the symbol $m$ to differentiate the different modes. Recall Fig.~\ref{fig:sh_table}. A pair of $n$ and $m$ then denotes a mode uniquely. $m$ always goes from $-n~\dots~n$. An $N$th-order ambisonic signal comprises all modes up to and including the ones for $n=N$. 

\begin{figure}[tb!]
    \centering
    \includegraphics[width=.48\columnwidth,trim=0in 0in 0in 0in, clip=true]{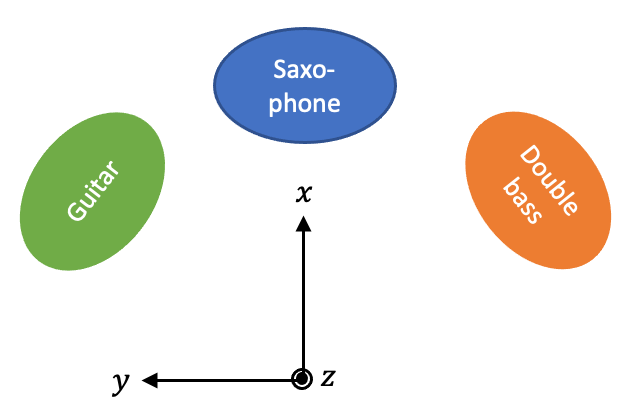}
    \caption{Sound scene encoded in the ambisonic signal from Tab.~\ref{tab:ambi_channels}. The $z$-axis points perpendicularly out of the plotting plane. The vantage point of the encoding is the coordinate origin.}
    \label{fig:recording_geometry}
    \vspace{3mm}
\end{figure}

Tab.~\ref{tab:ambi_channels} provides audio examples for the different channels of an example ambisonic signal that contains a scene composed of a saxophone, a guitar, and a double bass. The saxophone is located straight ahead of the vantage point, the guitar at \SI{45}{\degree} to the left, and the double bass at 45° to the right as illustrated in Fig.~\ref{fig:recording_geometry}.

\begin{table}[tb!]
\centering
\begin{tabular}{|c|c|l|}
\cline{1-3}
Channel & Mode $(n,m)$ & File name  \\ \hhline{|=|=|=|}
 1 & (0, 0) &  \texttt{channels/channel\_0\_0.wav} \\ \cline{1-3}
 2 & (1, -1)  & \texttt{channels/channel\_1\_-1.wav} \\ \cline{1-3}
 3 & (1, 0) & \texttt{channels/channel\_1\_0.wav}  \\ \cline{1-3}
 4 & (1, 1) & \texttt{channels/channel\_1\_1.wav}  \\ \cline{1-3}
 5 & (2, -2) & \texttt{channels/channel\_2\_-2.wav}  \\ \cline{1-3}
 6 & (2, -1) & \texttt{channels/channel\_2\_-1.wav}  \\ \cline{1-3}
 7 & (2, 0) & \texttt{channels/channel\_2\_0.wav}  \\ \cline{1-3}
 8 & (2, 1) & \texttt{channels/channel\_2\_1.wav}  \\ \cline{1-3}
 9 & (2, 2) & \texttt{channels/channel\_2\_2.wav}  \\ \cline{1-3}
\end{tabular}
\caption{Audio examples\footref{note2}\textsuperscript{,}\footref{note1}~of the individual channels of an ambisonic signal (saxophone + guitar + double bass, see Fig.~\ref{fig:recording_geometry}). The channels are sorted according to the ACN scheme. The ambisonic signal was created via a computer simulation.}
\label{tab:ambi_channels}
\end{table}
%

\footnotetext[2]{\label{note2}The audio examples can be listened to online at \url{https://www.ta.chalmers.se/research/audio-technology-group/audio-examples/introduction-to-ambisonics/} and can be downloaded from \url{https://doi.org/10.5281/zenodo.11060022}.}

\footnotetext[3]{\label{note1}The source signals are from \url{http://www.lam.jussieu.fr/Projets/index.php?page=AVAD-VR}.}

\setcounter{footnote}{3}

Recall from Fig.~\ref{fig:sh_table} that the mode $(n, m) = (0, 0)$ comprises the equivalent of an omnidirectional signal. You will hear all instruments equally loud in that channel. You might be noticing that the saxophone sounds a  quieter and more reverberant in the mode $(n, m) = (1, -1)$. This is because the direct sound of the saxophone happens to impinge from a direction in which the `directivity' of the mode has a null. As a consequence, the direct sound is suppressed (while the reverberation is not as it impinges from different directions). This is similar for the mode $(n, m) = (1, 0)$ in which the direct sound is suppressed for all three sound sources as the horizontal plane lies entirely in the null of the directivity. The saxophone is louder in the mode $(n, m) = (1,1)$ because that mode has a lobe that points directly at the saxophone. The contents of most of the other channels are very difficult to interpret when listening to them. This holds especially true for all modes $n > 1$. They simply sound like variants of the other channels.

We conclude that in some situations, it may be possible for some of the channels of an ambisonic signal to make sense of what one hears when listening to a given channel in isolation. But this will be very difficult in the general case. Ambisonic signals unfold their magic when the different channels are combined by means of mathematical operations. We will see this in the subsequent sections.

\section{Rendering (a.k.a.~Decoding) of Ambisonic Signals}\label{sec:reproduction}

Let us disregard for the time being where our ambisonic signal of interest might have originated from and first look at how it can be reproduced. Once we have covered that, it will be easier to get our heads around how to capture or create such a signal.

Ambisonic signals can be reproduced through loudspeaker arrays as well as binaurally whereby loudspeaker arrays are the original way. One also speaks of \emph{decoding} or \emph{rendering} the encoded scene.

\subsection{Loudspeaker Arrays}\label{sec:ls_reproduction}

Decoding ambisonic signals for a loudspeaker array means computing the signals with which a given set of loudspeakers need to be driven to make a listener experience the sound scene that the ambisonic signal represents. We will explain further below why decoding in real life is an intricate process in terms of the signal processing that is being carried out so that we have to rely on software tools (cf.~Sec.~\ref{sec:tools}). We want to first attempt an explanation that is simplifying but may help the reader develop an intuitive understanding of the process of decoding. 

\begin{figure}[b!]
    \centering
    \includegraphics[width=.4\columnwidth,trim=0in 1.2in .4in 1.1in, clip=true]{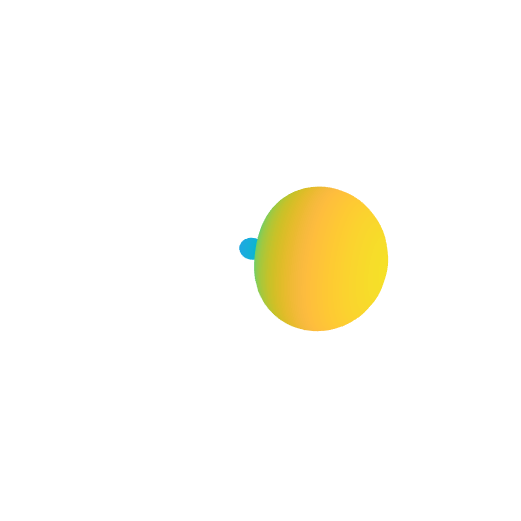}%
    \hspace{.5cm}\includegraphics[width=.4\columnwidth,trim=.5in 1.2in 0in 1.1in, clip=true]{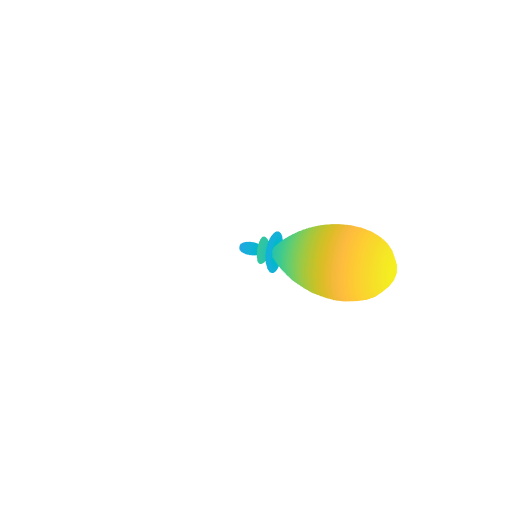}%
    \caption{The theoretical microphone directivities that arise if the channels of an ambisonic signal are added in a specific fashion as described in the text.}
    \label{fig:decoding_directivity}
\end{figure}

Recall Fig.~\ref{fig:sh_table}, which displays the directivity that a notional microphone needs to have to capture the signal that is in a given channel of an ambisonic signal. If we add the channels of an ambisonic signal that correspond to the modes $(0, 0)$ and $(0, -1)$, i.e., we add channels 1 and 2 if ACN is used, we obtain a new signal that represents the signal that would be captured by a notional microphone with the directivity that is depicted in Fig.~\ref{fig:decoding_directivity} (left): A cardioid pointing into positive $y$-direction (a slightly higher weight was put on the figure-of-eight directivity compared to the omnidirectional one in this figure). If all channels from Fig.~\ref{fig:sh_table} are added with a specific relative weighting (gain), then a new signal is obtained that represents the signal that is captured by a notional microphone with the directivity that is depicted in Fig.~\ref{fig:decoding_directivity} (right): A highly directional microphone with a main lobe pointing into positive $y$-direction. In other words, the microphone in Fig.~\ref{fig:decoding_directivity} (right) captures mostly only those components of the sound field that impinge from the direction into which the main lobe points. If we position a loudspeaker in that very direction and drive it with that very signal, the loudspeaker reproduces the corresponding sound field components with the correct propagation direction. If this is done similarly for a suitable set of directions all around the listener, then the sound scene that is represented by the ambisonic signal is reproduced as a whole. It is being \emph{decoded}. Note that directivity lobes like the ones from Fig.~\ref{fig:decoding_directivity} can be steered to arbitrary directions (adding, for example, two figure-of-eight directivities produces another figure-of-eight directivity oriented along a different axis and so forth).

\begin{figure}[tb]
    \centering
    \includegraphics[height=.27\columnwidth,trim=0in 0in 0in 0in, clip=true]{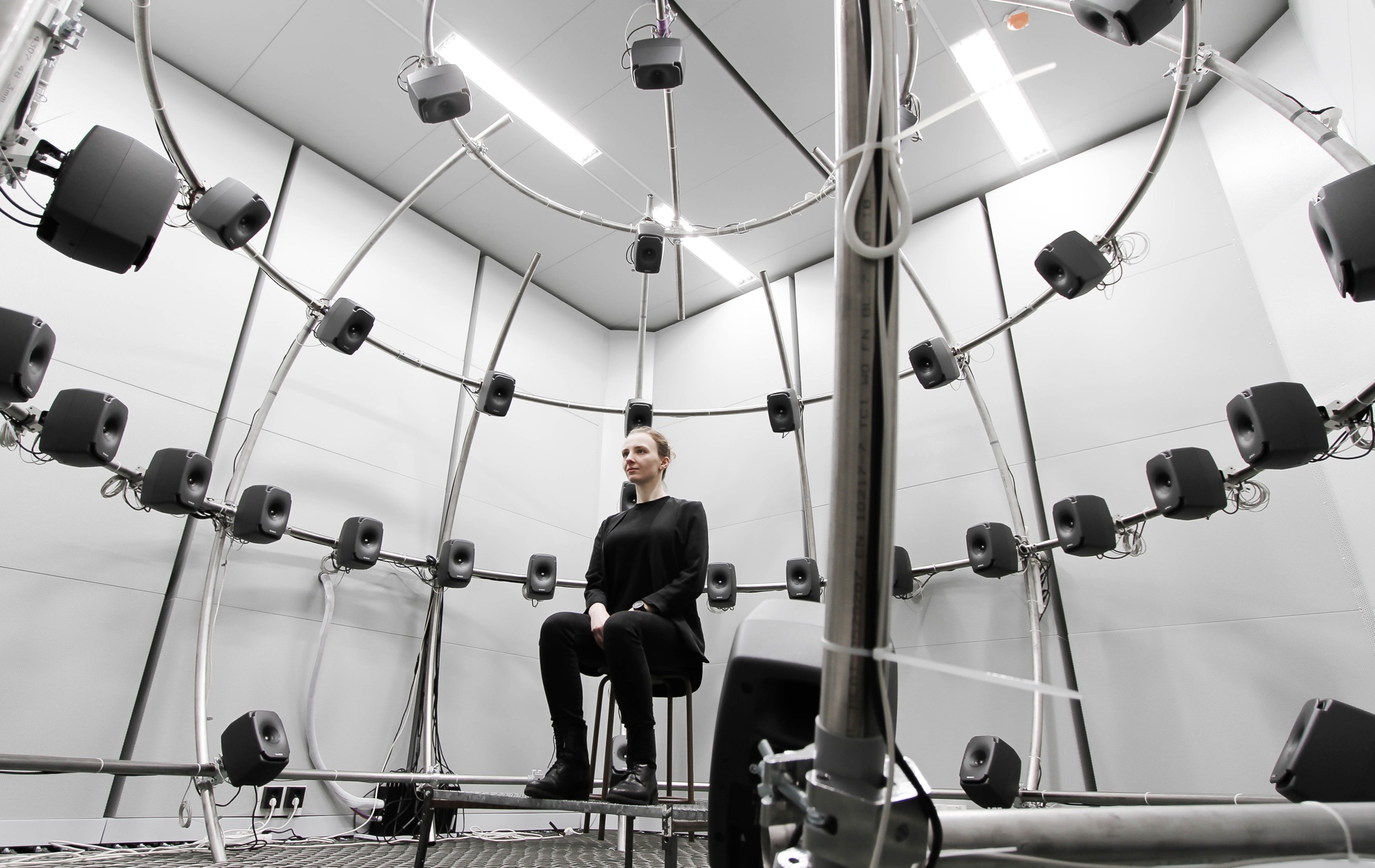}\hfill\includegraphics[height=.27\columnwidth,trim=0in 0in 0in 0in, clip=true]{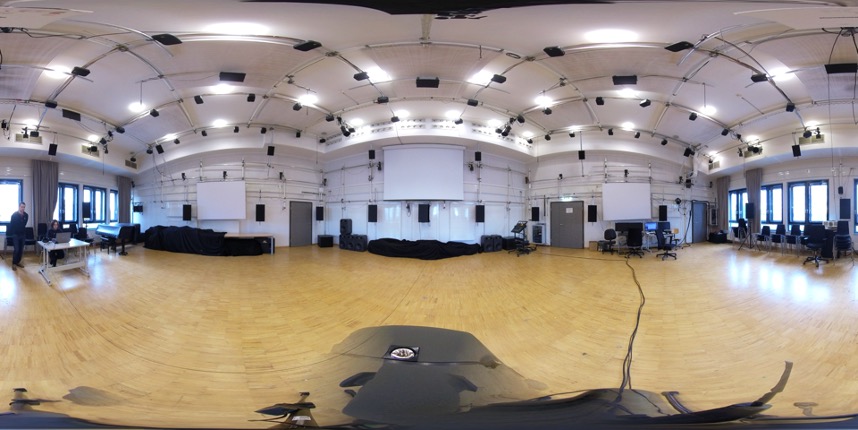}
    \caption{Photographs of ambisonic loudspeaker arrays. Left:  Room Wilska at Aalto University (45 loudspeakers on a sphere). Right: IEM Cube (25 loudspeakers on a dome plus 5 subwoofers). (Image credits: Beno\^it Alary, Franz Zotter)}  
    \label{fig:ambi_arrarys}
\end{figure}

Decoding works best if the loudspeakers are arranged on a sphere. Loudspeaker arrangements that deviate from spherical (such as ellipsoids and the like) can be used, too, in principle, whereby there is no standard solution for computing the loudspeaker signals in these cases, and the result will vary depending on the chosen decoding method. Horizontal-only arrangements such as circles of loudspeakers are possible, too. Refer to Fig.~\ref{fig:ambi_arrarys} for a few examples. 

As a general rule, the loudspeaker arrays need to enclose the listening area, and the more they deviate from a spherical or circular shape, the more limitations arise with respect to the reproduction accuracy. Reproduction of an $N$th-order ambisonic signal requires technically at least $(N+1)^2$ loudspeakers to fully render all information in the signal. Some of the recent decoders like \href{https://plugins.iem.at/docs/allradecoder/}{AllRAD} can work with fewer loudspeakers and with loudspeaker setup that are not homogeneous (for example, if there are no loudspeakers below the horizontal plane)~\citep{Zotter:allrad}. Circular loudspeaker arrays require in theory $2N+1$ loudspeakers but are not capable of reproducing elevation information.

\begin{figure}[tb]
    \centering
    \hfill\includegraphics[width=.33\columnwidth,trim=7in 1in 6in 0in, clip=true]{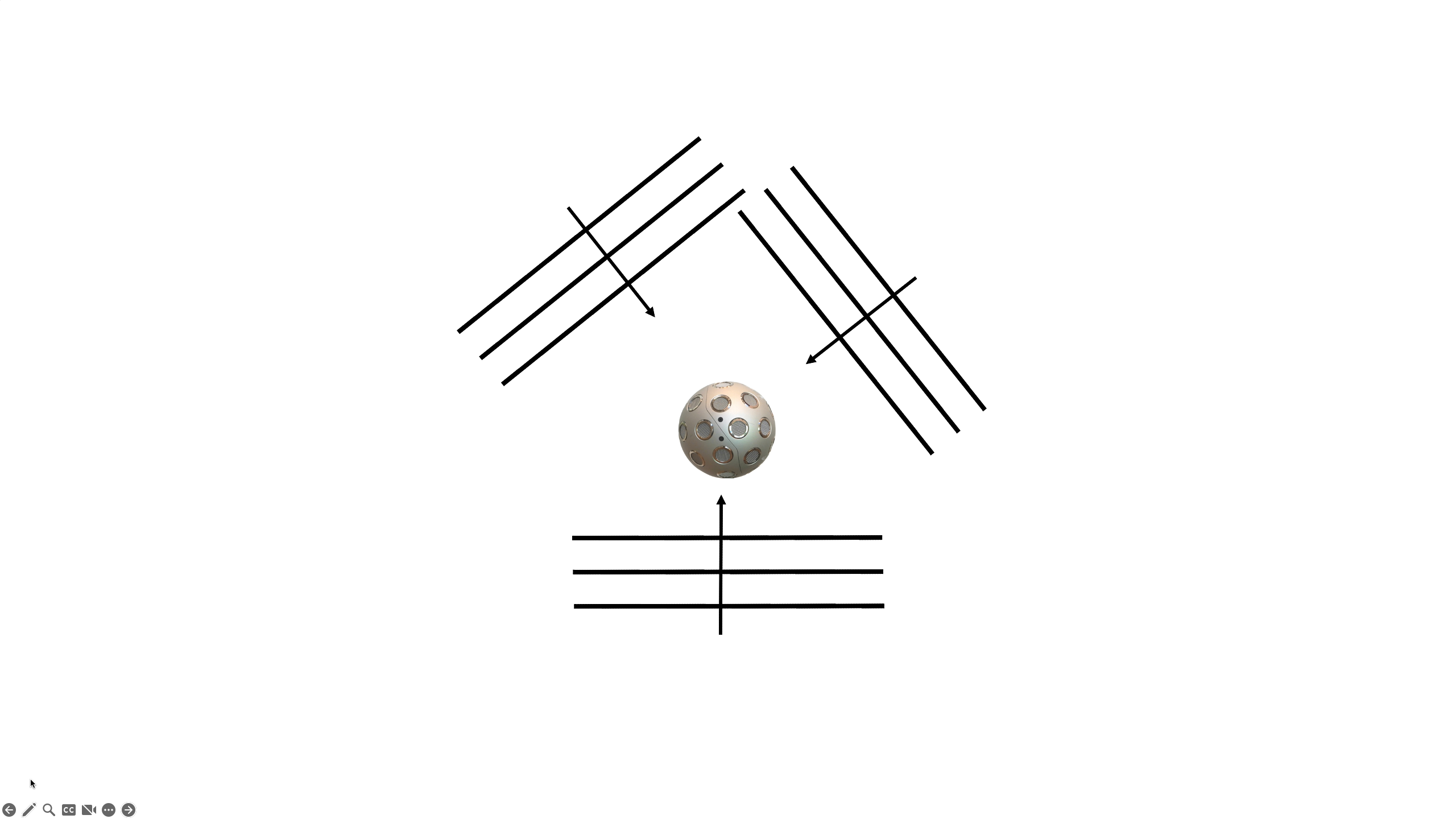}\hfill\includegraphics[width=.33\columnwidth,trim=7in 1in 6in 0in, clip=true]{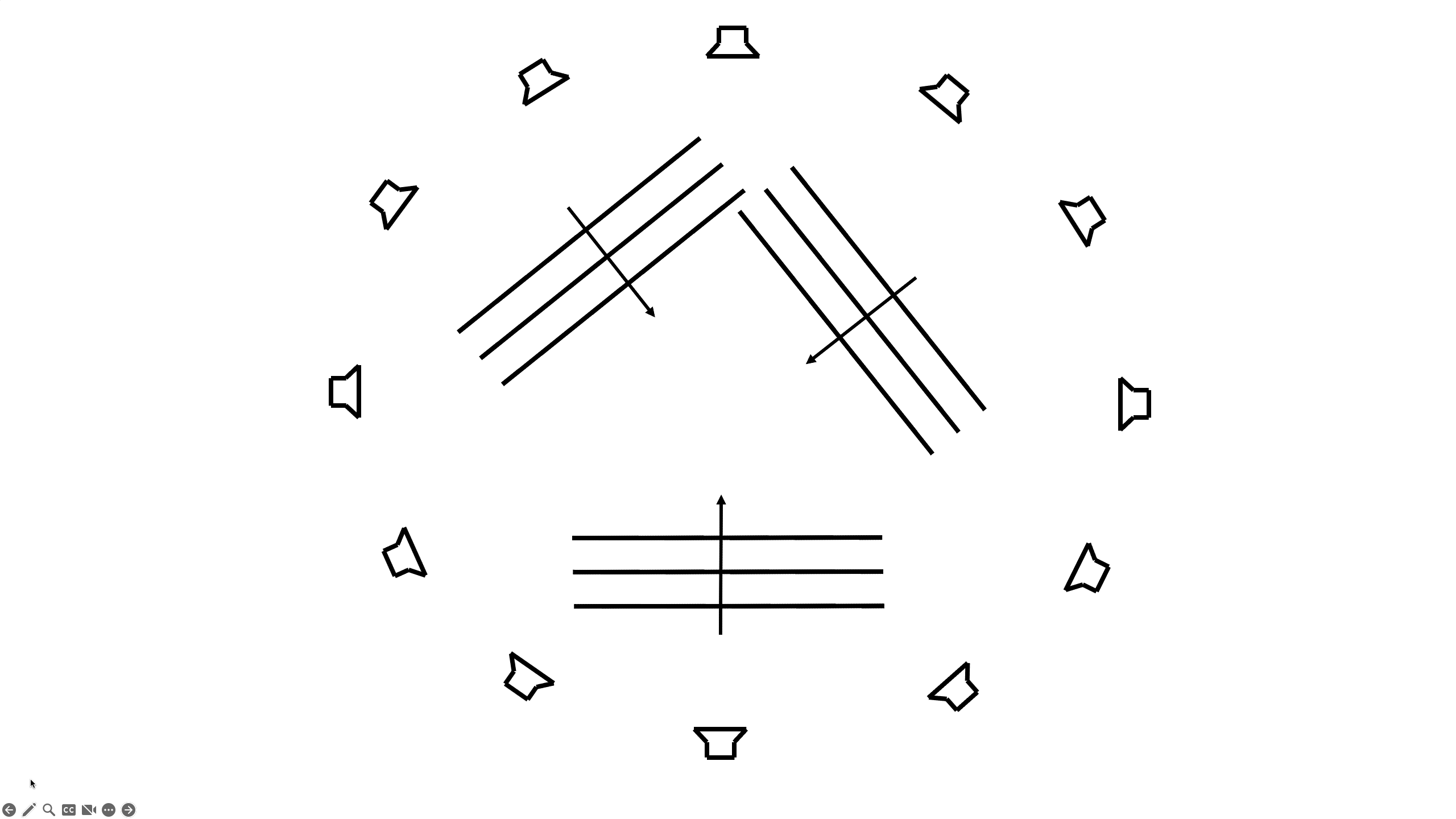} \includegraphics[width=.33\columnwidth,trim=7in 1in 6in 0in, clip=true]{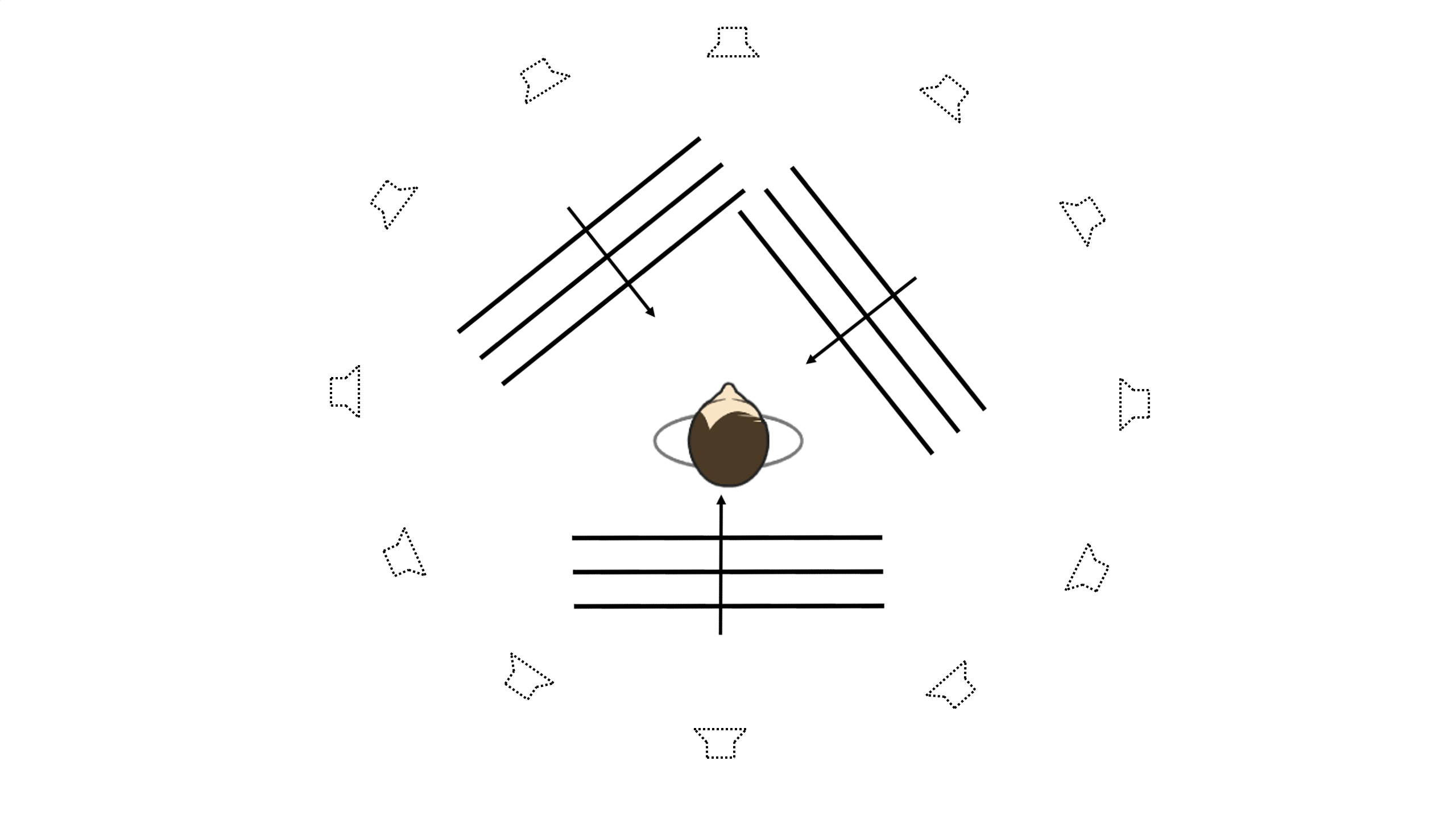}\hfill%
    \put(-445, 0){1)}\put(-310, 0){2)}\put(-155, 0){3)}
    \caption{Illustration of binaural rendering. Step 1: A microphone captures a sound field and encodes it into ambisonics. Step 2: A notional loudspeaker array recreates (decodes) the sound field that is encoded in the ambisonic signal. Step 3: Filtering each loudspeaker signal with the pair of HRTFs that correspond to the loudspeaker position creates the ear signals that the listener would have experienced in the original sound field. The loudspeaker array is 'virtualized'.}
    \label{fig:ambi_binaural_rendering}
\end{figure}

Given that an ambisonic signal is a representation of the physical structure of the captured sound field, it is theoretically possible to derive a mathematical formulation for the signals that a given loudspeaker array needs to be driven with so that the loudspeaker array physically reconstructs the encoded sound field. Such decoding could be considered the ideal reproduction of the encoded sound field and is illustrated in Fig.~\ref{fig:ambi_binaural_rendering} (middle). Things are unfortunately more complicated in real life. 

The main reason for this is that a practically feasible ambisonic signal represents the captured sound field correctly only in a small portion of space around the vantage point. For a 1st-order signal, this portion of space is smaller than a human head so that even the signals that arise the ears of a listener that is perfectly centered in the loudspeaker array are physically not correct. This is the case even for ambisonic signals with high orders. Actually, for the reproduced sound field to be physically correct over the entire frequency range of audible sound inside a portion of space that is large enough to comprise a human head, an ambisonic order higher than 30 would be required. Recall that this entails $31^2=961$ or more audio channels. Even then, the listener may not move outside of the central spot.

Obviously, a different strategy is required for gaining access to the information in the ambisonic signals. This led to the formulation of perceptual criteria based on which the decoders were designed~\citep{Gerzon:1992}. In other words, decoders were designed such that the reproduced sound scene \emph{sounds} as close as possible to the original one even though the ear signals that are evoked are different from the `true' ones. One of the strongest such criteria is that the direction of the flow of energy in the reproduced sound field should correspond to the direction of flow of energy in the original one. The psychoacoustic criteria are usually applied at higher frequencies and are blended with the criterion of physical accuracy at low frequencies. Decoder design is a delicate problem, and dedicated tools like~\citep{Heller:LAC2012} are available. Some plugins for digital audio workstations like\footnote{\url{https://plugins.iem.at/}, \url{https://leomccormack.github.io/sparta-site/}, \url{https://www.ambisonictoolkit.net/}} also comprise state-of-the-art decoders. A very popular such decoding approach is termed AllRAD and was originally presented in~\citep{Zotter:allrad}. A variety of other decoding methods are summarized in~\citep[Sec.~4.8-4.9]{Zotter:book2019}.

As with any other things, it is not such that a certain decoder is better than another. They are just different. And they do sound different. This opens opportunities as one may choose the decoder method according to the desired aesthetics of the scene that is to be reproduced.

%

\subsection{Headphones}

In order to understand binaural decoding of ambisonic signals, we need to recall the concept of head-related transfer functions (HRTFs). HRTFs represent the transfer path from a loudspeaker in an anechoic space to the two ears of the listener. In other words, if one filters a signal say, speech, with the pair of HRTFs that correspond to a given loudspeaker position, then the result is the signals that would arise at the listener's ears if the loudspeaker would have really played that speech signal. This may be interpreted as a virtual loudspeaker playing the speech signal.

HRTF sets are usually measured with loudspeakers on a spherical surface centered around the listener's head. This means that an HRTF set is nothing other than a virtual spherical loudspeaker array! We can simply decode any ambisonic signal on the virtual loudspeaker array, then filter the loudspeaker feeds with the HRTFs that correspond to the loudspeakers' positions and add up the signals. The result is the signals that would arise at the listener's left and right ears if the loudspeakers with which the HRTF set was measured would have reproduced the ambisonic scene. All we need to do then is play back the signal over headphones. Fig.~\ref{fig:ambi_binaural_rendering} summarizes the concept. There are some advanced mathematics that allow for virtually rotating the listener's head in the virtual loudspeaker array so that head tracking can be applied upon playback. 

In practise, binaural decoding is done slightly differently than loudspeaker decoding because significantly more virtual loudspeakers are usually available for binaural decoding than real loudspeakers in loudspeaker decoding. Also, the listener is guaranteed to be located in the center of the array in binaural decoding. This allows for a variety of optimizations that cannot be applied with loudspeaker decoding~\citep[Ch.~4.11]{Zotter:book2019}.

\section{Capture (a.k.a.~Encoding) of Ambisonic Content}\label{sec:encoding}

\begin{figure}[tb]
\centering
\begin{minipage}{.47\textwidth}
\vspace{1mm}
    \centering
    \includegraphics[width=.6\columnwidth,trim=0in 3in 0in 0in, clip=true]{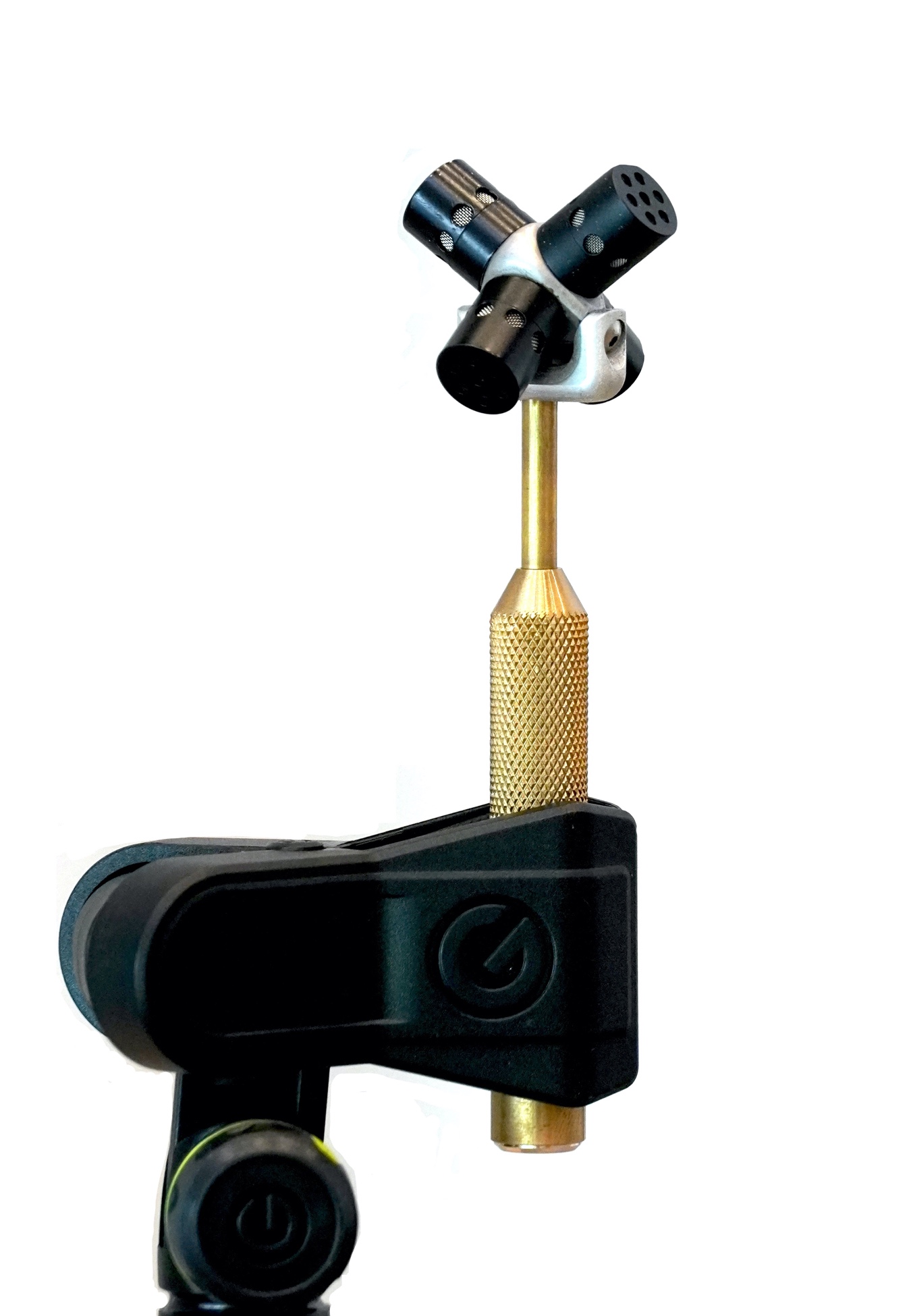}
    \caption{Photograph of a commercial tetrahedral microphone. It is composed of four nearly coincident outward-facing cardioid-like capsules. It can deliver a 1st-order ambisonic signal. The head with the microphone capsules has a diameter of approx.~\SI{4}{cm}. (Photograph: Jens Ahrens)}
    \label{fig:tretramic}
    \end{minipage}
\hfill
\begin{minipage}{.47\textwidth}
    \centering
    \includegraphics[width=1\columnwidth,trim=0in 0in 0in 0in, clip=true]{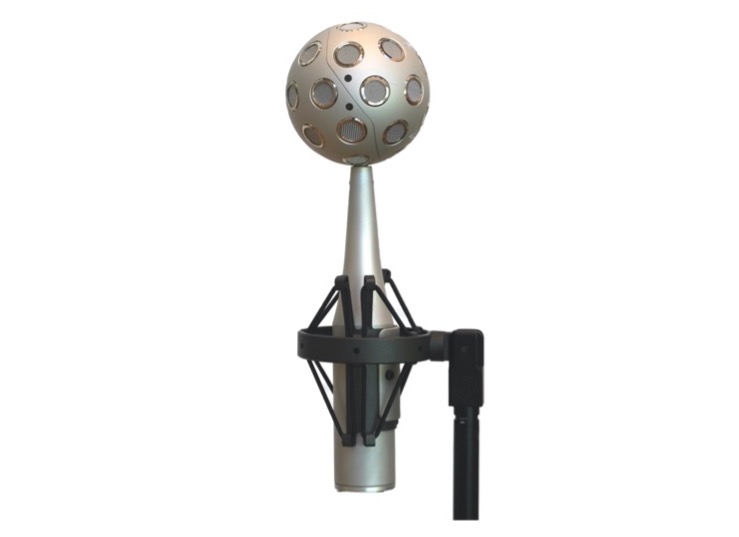}
    \caption{Photograph of the commercial spherical microphone em32 Eigenmike\textsuperscript{TM} with 32 channels and a diameter of \SI{8.4}{cm}. It can deliver a 4th-order ambisonic signal. (Photograph: Jens Ahrens)}
    \label{fig:eigen_mic}
\end{minipage}
\end{figure}

In the early years of ambisonics, the only way to obtain ambisonic signals was by recording them with a tetrahedral microphone array like the one depicted in Fig.~\ref{fig:tretramic}. Such an array comprises four microphone capsules that are positioned in the corners of a tetrahedron. The capsules have a cardioid-like directivity, and their main lobes point outwards in radial direction.  Adding the signals from the four microphones produces an omni-directional directivity that represents the first channel of the ambisonic signal (assuming ACN, first row of Fig.~\ref{fig:sh_table}). 
Adding and subtracting the signals in different combinations produces the virtual figure-of-eight directivities that correspond to the second, third, and fourth channels of the ambisonic signal (second row of Fig.~\ref{fig:sh_table}). One also speaks of the captured sound field being encoded into ambisonics. In other words, for this microphone array, the process of encoding the captured sound field consists simply in adding and subtracting microphone signals. The resulting four-channel ambisonic signal is termed \emph{B-format}, if the channels are ordered in a specific way (that is different from ACN, see Sec.~\ref{sec:terminology}).

By the way, the resulting virtual omni and figure-of-eight microphones can be combined to produce cardioid, hyper-cardioid, etc.~directivities the orientations of which can be steered electronically upon postprocessing simply by changing how the microphone signals are combined. This is not directly related to ambisonics but good to know, too.

There is a variety of microphone array designs that can capture ambisonic signals with an order higher than 1. The most established design is that of spherical microphone arrays. These comprise rigid spherical baffles that are of a size that is typically between that of a tennis ball and that of a bowling ball. See Fig.~\ref{fig:eigen_mic} for an example. Yet again, the rule is that a microphone array that is to deliver an ambisonic signal of order $N$ requires at least $(N+1)^2$ microphones. The 32-channel array from Fig.~\ref{fig:eigen_mic} produces a 4th-order signal. The mathematics and the signal processing that convert the raw microphone signals into ambisonics are very advanced, and you will have to trust the software tools that you are using (cf.~Sec.~\ref{sec:tools}). It is not possible for a user to barge in.

\begin{figure}[tb]
    \centering
    \includegraphics[width=.45\columnwidth,trim=0in 0in 0in 0in, clip=true]{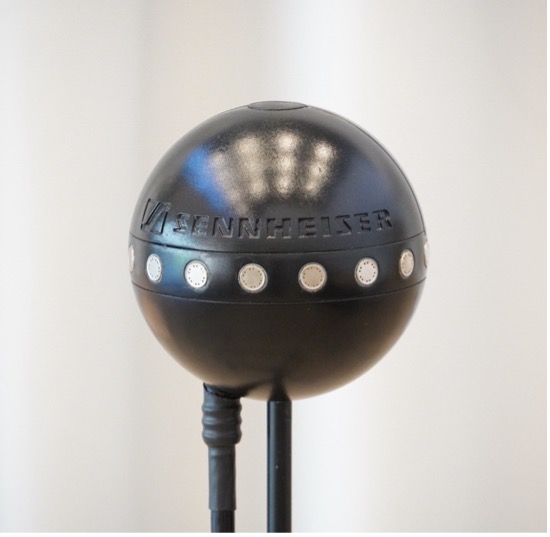}
    \caption{16-channel equatorial microphone array with a diameter of \SI{16}{cm} that produces 7th-order ambisonics. The hardware design is described in~\citep{Fiedler:2018}. Find a demo of it here\footref{note4}. (Photo: Jens Ahrens)}
    \label{fig:ema}
\end{figure}

As mentioned before, a variety of other microphone array designs exist for high-order ambisonic capture that are not as established (yet) as spherical microphone arrays. One example that we want to present here is the equatorial microphone array depicted in Fig.~\ref{fig:ema}. It uses also a spherical baffle, just like spherical microphone arrays, but its microphones are arranged exclusively along the equator of the baffle. This reduces ever so slightly its ability to capture information on the elevation of the sounds that it captured, but the advantage is that this design requires only $2N+1$ microphones to capture  $N$th-order ambisonics. The 16-channel equatorial array from Fig.~\ref{fig:ema} captures 7th-order ambisonics whereas the 32-channel spherical array from Fig.~\ref{fig:eigen_mic} captures 4th-order ambisonics (cf.~Tab.~\ref{tab:binaural_examples} in Sec.~\ref{sec:ambi_order} for audio audio examples of different orders).

It is also possible to encode virtual sound scenes. In a nutshell, this requires choosing a source position relative to the vantage point and imposing the directivities from Fig.~\ref{fig:sh_table} onto the source signal in software. This way, virtual sound sources can be arbitrarily positioned in space just like with panning-based reproduction methods. The ambisonic signal that represents a given virtual sound source is simply added to the ambisonic signal that represents the rest of the scene. It is technically possible to add ambisonic signals of different orders simply by adding the corresponding channels. This does not come without limitations mainly because renderers would optimize the rendering for the overall order of the ambisonic signal. This can lead to sub-optimal rendering for the lower-order components of the sound scene.

One may be tempted to assume that ambisonic microphone arrays will not be good for capturing diffuse reverberation given that all microphone capsules are almost coincident. In the encoding process, the signals from the individual microphone capsules are combined using signal processing, which may be interpreted as creating a meta-microphone that exhibits a directivity that is much more pronounced than what a single capsule can ever achieve. This very pronounced directivity counteracts the circumstance that the microphone capsules are located so close to each other so that it is indeed possible to obtain an ambisonic signal with very diffuse reverberation (which is equivalent to obtaining several channels of uncorrelated reverberation). This works better the higher the order of the ambisonic signal is.

\section{The Significance of the Order}\label{sec:ambi_order}

The order of an ambisonic signal (or an ambisonic capture or playback system for that matter) is a very fundamental parameter. As mentioned above, the number of loudspeakers in a playback setup or the number of capsules in a capture setup determine the maximum order that is supported. The actual cause of this limitation is buried deep in the physics of the problem and is beyond the scope of this tutorial. From a practical perspective, the rule of thumb is that an $N$th-order playback system can only reproduce the lowest~$N$ orders of the ambisonic signal that it is driven with even if that signal comprises also orders higher than $N$. No particular limitations arise if a loudspeaker array supports higher orders than what an ambisonic signal may comprise. Yet, many decoders are designed such that the result may not be optimal if the order of the content is not equal to the maximum order that the reproduction system supports.

\footnotetext[4]{\label{note4}Equatorial array demo: \url{https://youtu.be/5jAu47l2WaY}}

\setcounter{footnote}{4}

The higher the order of an ambisonic signal is, the higher is the accuracy with which it represents the physical structure of the captured sound field. What this means perceptually, i.e., in what way this changes what the sound scene sounds like, is difficult to say in a few lines~\citep{Barrett:2012}. Changing the order has different effects when decoding the scene on a loudspeaker array compared to when decoding it binaurally.  

When it comes to loudspeaker reproduction, a distinct effect of the ambisonic order is the size of the \emph{sweet spot}, i.e.~the preferred listening location. There is no rigid definition of what a sweet spot is. It is typically the position or the area in which perceptual properties like localization and timbre are optimal. How much these properties need to deviate from optimal so that one is declared to be outside of the sweet spot or sweet area is not defined. 

\begin{figure}[tb]
\hspace{1.5cm}%
\includegraphics[width=.35\columnwidth,trim=0in 0in 0in 0in, clip=true]{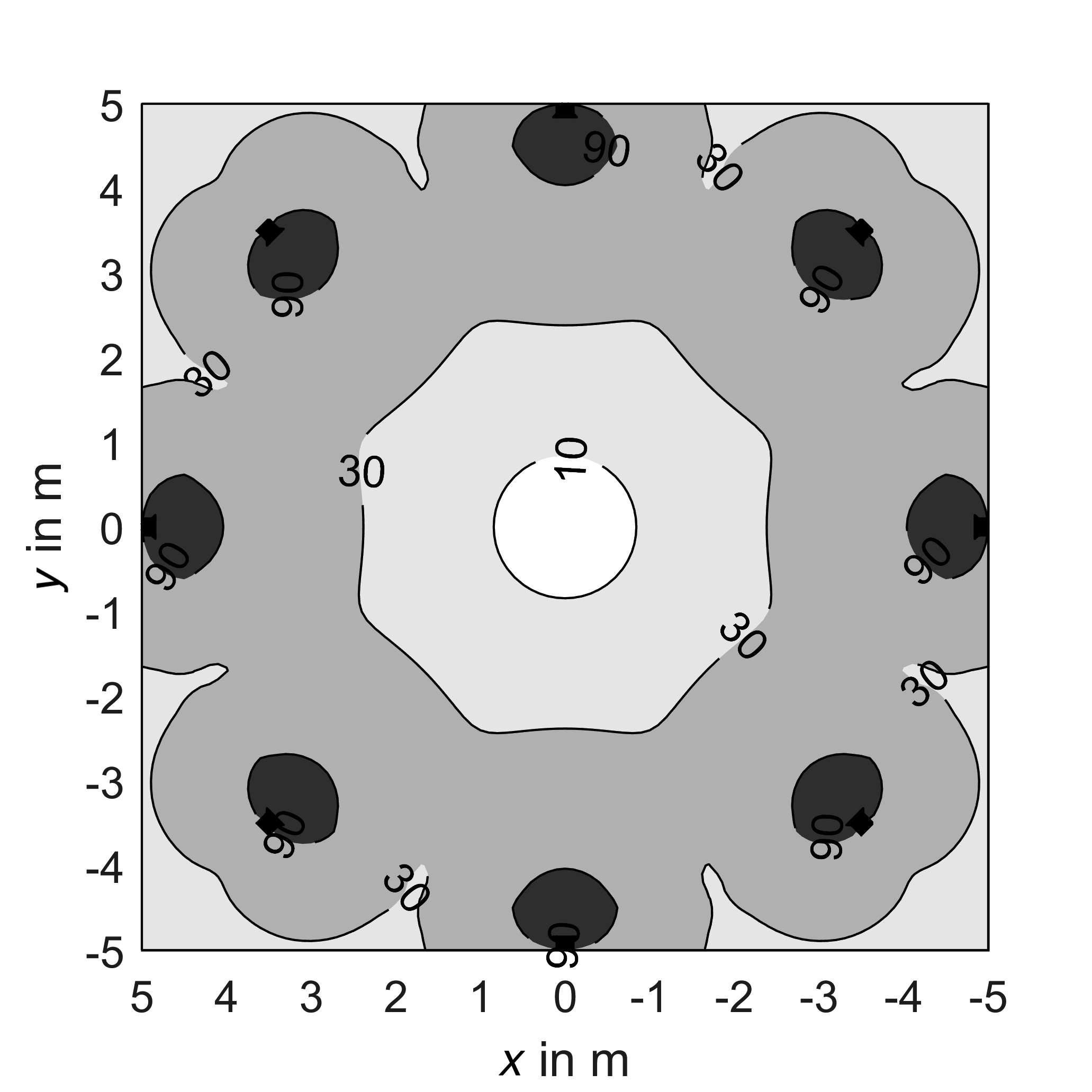}\hspace{1.5cm}%
\includegraphics[width=.35\columnwidth,trim=0in 0in 0in 0in, clip=true]{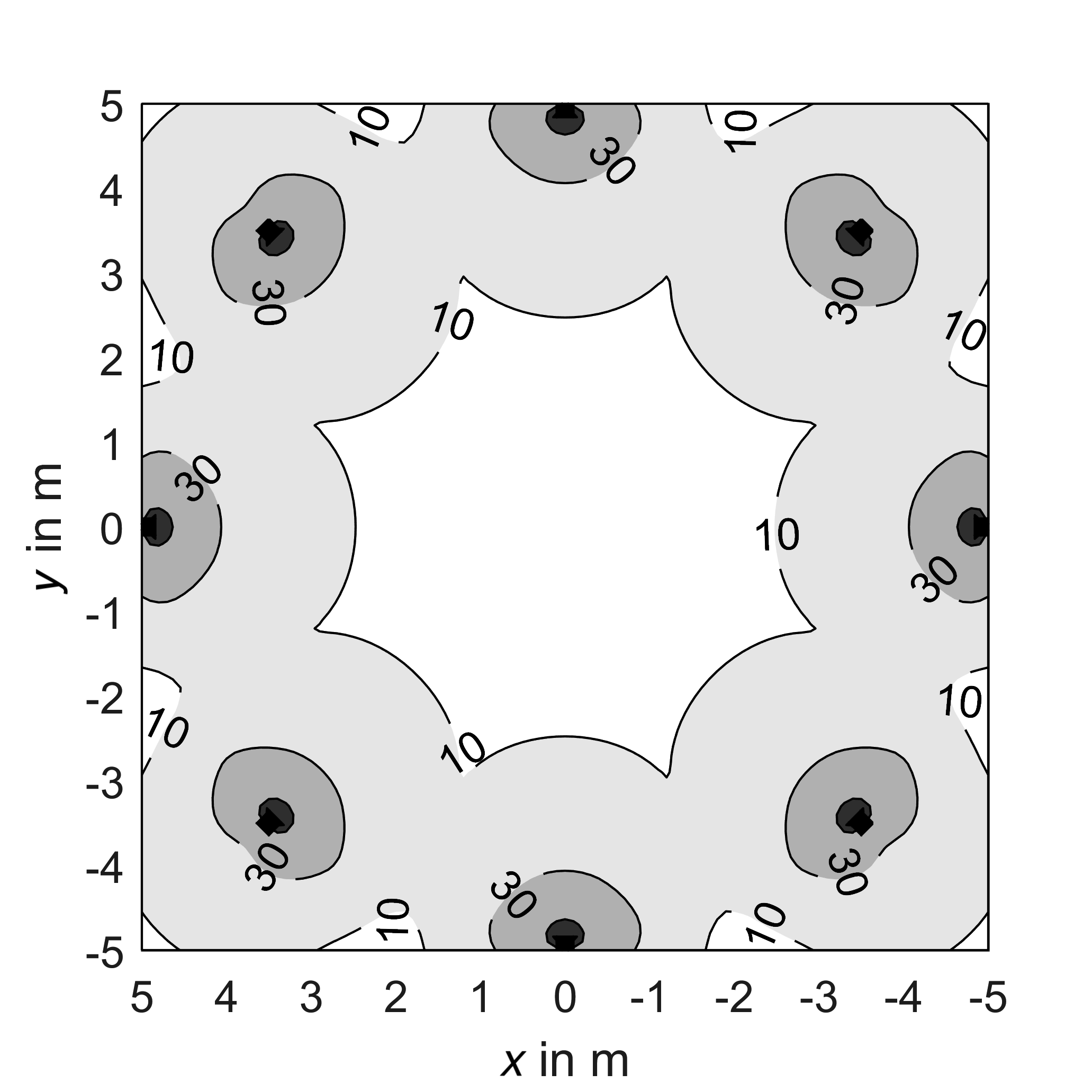}
\caption{Illustration of the localization uncertainty in degrees in an 8-channel circular loudspeaker array for 1st order (left) and 3rd order (right) reproduction. Images from~\citep[Fig.~4.7, CC-by-4.0]{Zotter:book2019}.}
\label{fig:sweet_spot}
\end{figure}

\begin{figure}[tb]
    \centering
    \includegraphics[width=.9\columnwidth,trim=0in 0in 0in 0in, clip=true]{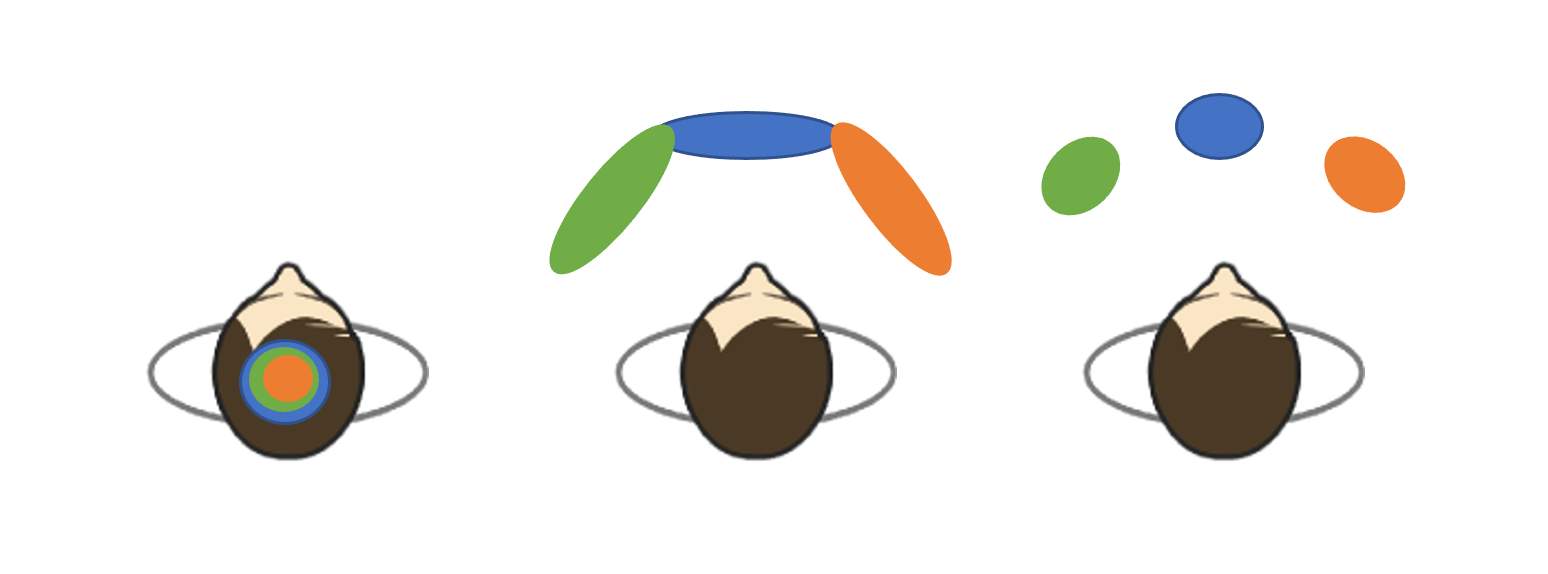}
    \caption{Illustration of the effect of the ambisonic order on locatedness. Left: 0th-order signal. Since a 0th-order signal does not contain any spatial information, all sound sources will appear at the exact same position, usually in the middle of the head. Middle: Low-order signal, say, 1st--3rd order. The sound sources in the scene are localizable, but their location might not be ultimately easy to determine, and some fluctuations arise when moving the head. Right: High-order signal, say, 4th order and above. The sound sources are clearly and stably localizable, and their spatial extent is compact.}
    \label{fig:sh_order}
\end{figure}

Fig.~\ref{fig:sweet_spot} depicts the predicted accuracy of the direction of the energy flow in the reproduced sound field for a circular 8-channel loudspeaker array. This accuracy of the direction of the energy flow is closely related to the uncertainty with which virtual sound sources are localized. We may therefore interpret the data from Fig.~\ref{fig:sweet_spot} as the localization uncertainty. The authors of~\citep{Zotter:book2019} suggest that all locations where the uncertainty is lower than \SI{30}{\degree} are the sweet area. For the 1st-order reproduction depicted in Fig.~\ref{fig:sweet_spot} (left), the sweet area extends from the center of the array to half-way to the loudspeakers; for the 3rd-order reproduction in Fig.~\ref{fig:sweet_spot} (right), the sweet area is almost as large as the entire area inside the loudspeaker array. The sweet area is accordingly smaller if one chooses a stricter requirement for the tolerable localization uncertainty. Either way, the size of the sweet area grows with the ambisonic order. Along with lower localization uncertainty at higher orders comes usually a sharpening of the localization meaning that the virtual sources are more clearly localizable and less blurry than in low-order reproduction. Fig.~\ref{fig:sh_order} illustrates this for binaural reproduction. Fig.~\ref{fig:sh_order} (middle) and (right) apply to loudspeaker reproduction, too, whereas 0th-order reproduction with loudspeakers means simply that all loudspeaker play the same signal.

Localization and related attributes like locatedness (i.e.~how clearly the source's position is perceived) are only one part of the whole. A secondary effect of higher spatial resolution in higher-order ambisonics is the its ability to produce lower interaural coherence for the listener, which correlates with higher perceived diffuseness and related attributes like higher perceived spatial depth~\citep[Sec.~1.5]{Zotter:book2019}.

We would like to illustrate the significance of the order on localization for binaural reproduction through listening to a few examples. See Tab.~\ref{tab:binaural_examples}. The listener is guaranteed to be in the sweet spot of the reproduction so that mere localization accuracy is not an issue. The timbre and the locatedness of the virtual sources are two of the affected perceptual properties in this case. The differences between different orders tend to be more obvious when head tracking is applied, which is unfortunately not possible here. The sound sources that you will hear will be circling around your head instead to achieve similar spatial dynamics like with head tracking.

In either case, binaural and loudspeaker-based, the output from the decoding is not perceptually convincing right away, and the signals need to be equalized. Tab.~\ref{tab:binaural_examples} presents both raw (unequalized) and equalized examples. This equalization involves complicated signal processing, and it is performed differently with loudspeaker-based reproduction compared to binaural reproduction. For loudspeaker-based reproduction, no standard procedure exists, and each decoding tool does the equalization slightly differently.

In binaural reproduction, the situation is somewhat more tangible. Here, the equalization equalizes the content of the scene direction-dependent. In other words, a source that is located on the right-hand side is equalized differently compared to a source that is located straight ahead. It is unfortunately not easy to demonstrate this without applying head tracking. Turning the head makes it clear that unequalized binaural decodings comprise direction-dependent coloration, which is mitigated by the equalization.

\begin{table}[tb]
\centering
\begin{tabular}{|c|l|l|}
\cline{1-3}
Order $N$ & Unequalized & Equalized  \\ \hhline{|=|=|=|}
 0 & \texttt{orders/binaural\_N0.wav} & \texttt{orders/binaural\_N0\_eq.wav} \\ \cline{1-3}
 1 & \texttt{orders/binaural\_N1.wav} & \texttt{orders/binaural\_N1\_eq.wav} \\ \cline{1-3}
 5 &  \texttt{orders/binaural\_N5.wav} & \texttt{orders/binaural\_N5\_eq.wav} \\ \cline{1-3}
 7 &  \texttt{orders/binaural\_N7.wav} & \texttt{orders/binaural\_N7\_eq.wav} \\ \cline{1-3}
\end{tabular}
\caption{Computer-generated binaural audio examples of different maximum orders $N$. The source signals are same like in Fig.~\ref{fig:recording_geometry} and Tab.~\ref{tab:ambi_channels} (from\footref{note1}) with the difference that all instruments are co-located. The column \emph{Unequalized} comprises the raw result, i.e., the binaural without any enhancement. The column \emph{Equalized} comprises the binaural signals with MagLS equalization applied (which SPARTA and the IEM Plugins, for example, provide; cf.~Sec.~\ref{sec:tools}). Focus particularly on how well the low-frequency content is localizable at the different orders and how smooth the motion of the sound sources is.}
\label{tab:binaural_examples}
\end{table}

Another practically relevant aspect that should not be overlooked is the data volume. A 7th-order ambisonic signal, for example, comprises $(7+1)^2 = 64$ channels. This means that we are talking file sizes of hundreds of megabytes even for moderate durations. 

In summary, it is difficult to state what order will be required in what situation. There are absolutely situations in which 1st order will be sufficient whereas others may benefit from a higher order. Applications that require orders higher than, say, 7 are rare in practice. 

\section{Manipulation of Ambisonic Representations}

As there is a strict mathematical relation between the contents of each of the channels, one should never manipulate a single channel or subset of channels of an ambisonic signal. The channels then simply lose their mathematical relation, and reproducing them can sound very different from what one would have expected. In other words, if one intends to fade out an ambisonic signal, for example, the exact same fade-out needs to be applied to all channels. Similarly, if one intends to apply an equalizer to an ambisonic signal, the exact same equalization needs to be applied to all of the channels~\citep[Sec.~5.3-5.4]{Zotter:book2019}. We emphasize here that being aware of this is of utmost importance. Dynamic compression of the sound scene as a whole is best performed with the $(n,m)=(0,0)$ channel, i.e.~the first channel of the ambisonic signal in ANC, as the control signal as this one comprises the signals from all sound sources in the scene. 

It is possible to leverage the mathematical relation between the channels and perform spatial transformations that are either not possible or very difficult to achieve with other spatial audio representations. As a user, we again have to rely on software tools for this as the mathematical operations that need to applied to the ambisonic signal can be fairly complicated. 

The most prominent examples for spatial manipulations that are possible in ambisonics are~\citep{Kronlachner:2014}:

\begin{itemize}
    \item Rotation of the entire sound scene
    \item Direction-dependent amplification or attenuation
    \item Mirroring (front-back, left-right, top-bottom)
    \item Directional warping (for example, squashing the content spatially towards a given direction)
\end{itemize}

The effect of all of the above manipulations tends to be stronger for signals with higher order except for rotation, which works always equally well. As mentioned, global manipulations like equalization, compression, fade-in/fade-out etc.~are possible so long as they are applied to all channels in the exact same manner. The by far most important of the spatial manipulations is rotation of the sound scene as a whole. Such rotations can be performed about an arbitrary axis, and they are lossless -- like many of the other spatial transformations. This means that if one rotates a scene by a given angle and rotates it back afterwards, one obtains exactly what one started with. Rotations are so important because they enable head tracking during binaural playback. One then speaks of 3-degrees-of-freedom (3~DoF) reproduction as rotations about all three Cartesian axes are possible. 

Direction-specific dynamic compression is an effect that is a hybrid between conventional audio processing and ambisonic processing: There are mathematical operations that allow for cutting out an angular segment of a sound scene, for example an angular segment that comprises, say, a singing voice that is to be dynamically compressed. The compression in then applied to the angular segment, which is added back into the sound scene afterwards. 

Note that translation of the vantage point (or equivalently of the listener in binaural reproduction) is generally not possible. In other words, the user cannot move around in the captured scene in binaural reproduction. A few methods have been proposed in the scientific literature for this, but all are still in an experimental state.

\section{Some Terminology}\label{sec:terminology}

The ambisonic community tends to use a terminology that can be very specific to ambisonics and the meaning of which cannot be inferred from other branches of spatial audio. We provide an overview of the most important terms in the following.\\

\begin{longtable}{ l l } 
Encoding & \parbox[t]{4.7in}{Process of converting the signals from the microphones of a microphone array to an ambisonic signal as well as conversion of the input signal of a virtual sound source and its metadata (such as its position in space) to an ambisonic signal\\}\\[4ex]
Decoding & \parbox[t]{4.7in}{Process of converting ambisonic signals into loudspeaker or headphone feeds}\\[1.5ex] 
ACN & \parbox[t]{4.7in}{Ambisonic channel number (the definition of which ambisonic mode to store in which channel of an ambisonic signal, see Fig.~\ref{fig:sh_table})}\\[4ex]
SID & \parbox[t]{4.7in}{Single Index Designation: A sorting scheme for the channels in an ambisonic signal that is different from ACN}\\[4ex] 
N3D, SN3D & \parbox[t]{4.7in}{3D normalisation and Schmidt semi-normalisation: Two different ways of normalizing the channels in an ambisonic signal relative to each other.}\\[4ex]
FuMa & \parbox[t]{4.7in}{Furse-Malham: Another sorting and normalization scheme for the channels in an ambisonic signal. See `B-format' below.}\\[4ex]
A-format & \parbox[t]{4.7in}{The raw microphone signals from a traditional 1st-order tetrahedral microphone array like the one depicted in Fig.~\ref{fig:tretramic}. Sometimes also used with arrays of order higher than~1.\\}\\[4ex] 
B-format & \parbox[t]{4.7in}{Traditional B-format refers to an ambisonic signal of 1st order (it comprises therefore four channels). The channels are termed W, X, Y, and Z. They correspond to the 1st, 4th, 2nd, and 3rd channel in ACN (ACN $=$ WYZX). FuMa is the extension of traditional B-format to higher orders. The signal format is then sometimes also simply referred to as B-format even if it is of order higher than 1 (and FuMa is sometimes also used with 1st-order signals).\\}\\[4ex] 
AMB & \parbox[t]{4.7in}{A file format for storing ambisonic signals. It is based on Microsoft's WAV\textsuperscript{TM} format and uses Furse-Malham channel ordering and normalisation. It is limited to 3rd order. File name extension: \texttt{.amb}\\}\\[4ex]
AmbiX & \parbox[t]{4.7in}{A file format for storing ambisonic signals. It is based on Apple's Core Audio Format\textsuperscript{TM} and uses ACN channel ordering with SN3D normalisation. File name extension: \texttt{.caf}}\\[4ex]
\end{longtable}

It is important that the settings of a software plugin are chosen correctly for the given ambisonic signal that is being processed, particularly regarding the channel order and the normalization.

At the time of writing, it seems to be uncommon to store ambisonic signals in \texttt{.amb} or \texttt{.caf}, and there are very few digital audio work stations that are able to open such files, if at all. Many times, \texttt{.wav} files are used, and the information on the channel order and the normalization is stored separately.

\section{Software Tools for Ambisonics}\label{sec:tools}

The circumstance that modern ambisonics received significant attention in the academic community before establishing in the industry contributed to the fact that a significant amount of free software tools are available some of which can have very advanced functionality. We provide a summary of some of the popular plugins in alphabetical order below. All of them provide basic functionality like encoding and decoding in binaural and for loudspeakers. We mention only examples for functionality that goes beyond this. As always, such a list can never be exhaustive, and there are plenty of other great ambisonic tools available.\\[2ex]

\begin{longtable}{ l l } 

\href{https://www.matthiaskronlachner.com/?p=2015}{ambiX} & Spatial manipulation\\ 
\href{https://harpex.net/}{Harpex} & Surround decoding and spatial upmixing \\ 
\href{https://plugins.iem.at/}{IEM Plugin Suite} & Vast set of plugins incl.~directional compressor, reverb, and other  \\
  & effects, AllRADecoder \\
\href{https://www.matthiaskronlachner.com/}{mcfx} & Not stricly ambisonic tools but very useful multichannel effects  \\
\href{https://www.blueripplesound.com/pro_audio}{O3A Core Plugin Library} & Broad set of functionalities \\ 
\href{https://forum.ircam.fr/projects/detail/panoramix/}{Panoramix} & Other spatialization methods and a variety of effects \\ 
\href{https://leomccormack.github.io/sparta-site/}{SPARTA} & Vast set of plugins incl.~a variety of effects incl.~source spread \\ 
\href{https://b-com.com/en/process/spatial-audio-family}{Spatial Audio family} & Encoding of recordings in multichannel formats like 5.1 \\ 
\end{longtable}

\section*{Acknowledgements}

This work received funding under the Erasmus+ programme of the European Commission under grant 2022-1-PL01-KA220-VET-000085305.

\bibliographystyle{plainnat}
\bibliography{references}

\end{document}